# Energy efficient ethanol concentration method for scalable CO$_2$ electrolysis

Magda H. Barecka,*[a,b,c], Pritika DS Dameni[c,d], Marsha Zakir Muhamad [c], Joel W. Ager *[e,f,g] and Alexei A. Lapkin [c,h]

[a.] Department of Chemical Engineering, Northeastern University, 360 Huntington Avenue, 02215 Boston, USA; email: m.barecka@northeastern.edu

[b.] Department of Chemistry and Chemical Biology, Northeastern University, 360 Huntington Avenue, 02215 Boston, USA

[c.] Cambridge Centre for Advanced Research and Education in Singapore, CARES Ltd. 1 CREATE Way, CREATE Tower #05-05, 138602 Singapore

[d.] Nanyang Technological University, Department of Mechanical Engineering, School of Materials Science and Engineering, 50 Nanyang Ave, 639798 Singapore

[e.] Berkeley Educational Alliance for Research in Singapore (BEARS), Ltd., 1 CREATE Way, 138602, Singapore

[f.] Department of Materials Science and Engineering, University of California at Berkeley, Berkeley, California 94720, USA

[g.] Materials Sciences Division, Lawrence Berkeley National Laboratory, Berkeley, California 94720, United States.

[h.] Department of Chemical Engineering and Biotechnology, University of Cambridge, Cambridge CB3 0AS, UK

## Abstract

Vacuum membrane distillation efficiently concentrates dilute ethanol streams produced by CO$_2$ electrolysis (CO2R), yielding up to ~40 wt.% ethanol in pure water. Our results allow for a more precise estimation of energy inputs to the separation processes and for the reformulation of CO2R technology development goals.

## TOC

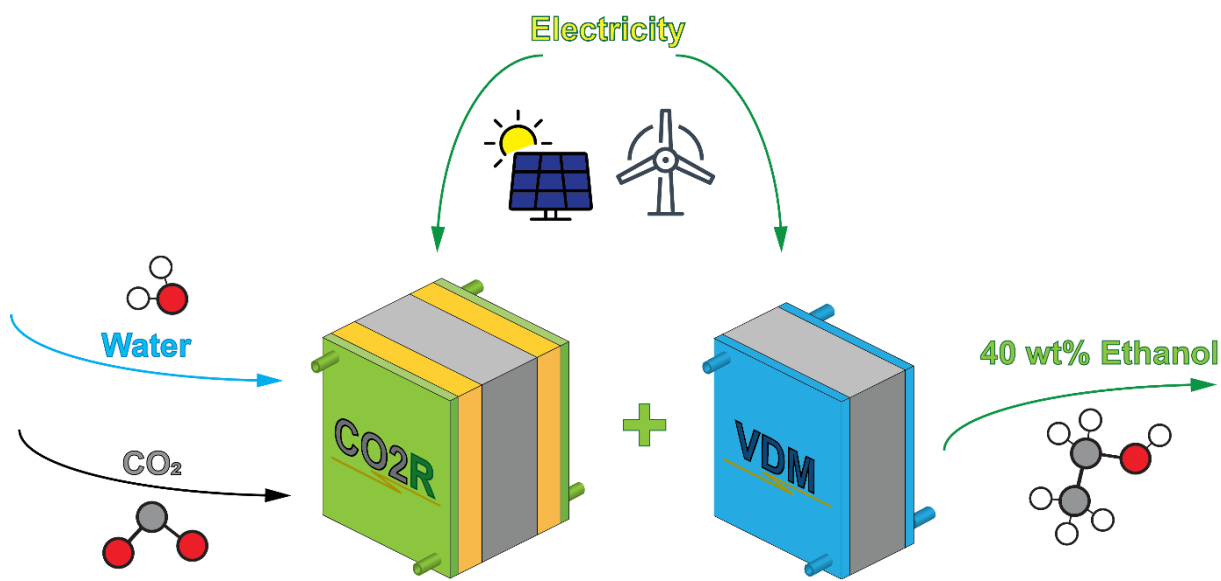

**Main Text**

Electrochemical conversion of carbon dioxide (CO2R) is a promising alternative for bulk chemicals production from fossil fuels[1,2]. If the CO2R process uses biogenic carbon dioxide ($CO_2$) and is powered by renewable energy, CO2R has the potential to yield indispensable chemical products with a minimum carbon footprint[3]. Among more than 15 compounds reported as potential products of CO2R[4], ethanol is highly relevant to the chemical industry, due to its use as fuel additive[5] and as a convenient precursor to production of non-fossil feedstocks derived ethylene[6]. However, it has been challenging to deploy CO2R to obtain ethanol streams of industrially relevant purity[7]; this communication showcases a first example of an economically viable approach to yield pure ethanol streams from CO2R.

During CO2R, ethanol product is accumulated in the liquid catholyte being continuously recirculated through the electrochemical reactor, whether a high-current gas diffusion electrode or an H-type cells are being used (**Fig. 1a**). The chemical and physical properties of the catholyte (e.g. pH, viscosity) are crucial to the selective and efficient electrolysis of $CO_2$, as the electrolyte layer adjacent to the CO2R catalyst belongs to the triple phase gas-solid-liquid boundary for the electrochemical reaction[8] (zoom in Fig. 1a). Thus, it is desirable to maintain a stable, controllable composition of the catholyte during CO2R experiments. High accumulation of any CO2R liquid product is not desirable as it also triggers crossover through the ion exchange membrane to the anode compartment, where ethanol is oxidized to acetic acid[9]. Thus, the concentration of ethanol produced by CO2R is frequently below 4 wt.%[10,11]; reports targeting production of higher ethanol concentration observe both increased crossover and limited stability[12].

While this does not influence the lab-scale characterization of new CO2R catalysts, it is a significant bottleneck towards the large-scale process deployment[13]: ethanol needs to be separated to allow for catholyte recycling, and the separation of highly diluted stream can require an energy input exceeding ethanol's heat of combustion[7]. Techno-economic analyses (TEA) of CO2R technologies point out that electrocatalytic reactors must yield min. 10 wt. % ethanol to be considered scalable[7,14,15]. The difficulties in reaching these goals made $CO_2$-to-ethanol less frequently investigated in terms of process scalability[16,17]. Thus, we sought to develop an energy efficient ethanol separation that will allow to leverage the progress in CO2R field to yield scalable $CO_2$-to-ethanol processes.

We first sought to understand the thermodynamic behaviour of liquids most frequently used as CO2R catholytes (e.g. potassium hydroxide solutions (KOH), in concentrations between 1-7 M[1,18]). Due to chemical absorption of $CO_2$ into KOH during CO2R, the catholyte turns into a mixture of KOH and salts: potassium bicarbonate ($KHCO_3$) and carbonate ($K_2CO_3$). While TEAs of CO2R processes typically quantify the energy inputs into ethanol separation for the ethanol-water binary system, from a thermodynamic standpoint, the separation of ethanol from catholyte solutions will be significantly different due to the salting-out effect[19,20]. When KOH is added to water, electrolyte's solvation decreases water and ethanol miscibility by reducing hydrogen bonding between water and ethanol.

To quantify the importance of salting-out in separation of ethanol from CO2R catholytes, we performed a series of headspace gas chromatography experiments. We quantified the amount of ethanol in the gas phase over solutions of 3 wt. % ethanol in different liquids: DI water, KOH in water, carbonate salts in water (**Fig. S1 a-b, raw data in Table S1**), at the same experimental conditions. The effect of salting-out is surprisingly high - the concentration of ethanol in the vapor phase above 7M KOH is six times higher than that above pure water, suggesting that a vapor-liquid equilibrium-based approach (distillation) can yield an efficient separation method.

Considering the design constraints of the entire CO2R-based chemical plant, we proposed to drive distillation by vacuum instead of thermal heat. This allows to avoid the need to cool down the catholyte after the separation step, before recycling it back to a CO2R reactor operating at room temperature. To increase the separation efficiency, we sought to enhance the mass transfer area using porous, hydrophobic polytetrafluoroethylene (PTFE) and polyvinylidene fluoride (PVDF) membranes which discriminate against water in the neighbourhood of the membrane and are compatible with the highly alkaline CO2R environment (pH ~14). Furthermore, as PVDF is also oleophilic, it further supports the concentration of ethanol close to the evaporation interface.

To evaluate vacuum membrane distillation (VMD) as a technique for CO2R liquid products separation we performed a series of experiments using a stainless-steel membrane contactor, vacuum pump and a control system embedded in a laboratory rotary evaporator. High-performance liquid chromatography (HPLC) was used for ethanol quantification (**Fig 1b, S2 and SI Experimental procedure**). We deployed membranes with different pore sizes and investigated the separation performance under a range of ethanol concentrations which can be obtained from $CO_2$ electrolysis (0.5 – 3 wt.% in 3.5 M KOH, **Table S2**). Across all studied conditions, the use of VMD allowed to substantially concentrate dilute ethanol in one step of separation and obtain pure ethanol product of concentration up to 39.5 wt.%, with the PVDF membrane yielding a higher ethanol concentration as it is more oleophilic (**Fig. 1c**).

Translating our results into the assessment of the energy input necessary for ethanol separation (**SI, Process Model**), we demonstrated that the use of VMD allows to recover dilute ethanol (0.5 – 3 wt. %) at a minimum energy input of 0.2 – 0.7 MJ/mol, which is comparable or lower than ethanol' heat of combustion[7] (1.4 MJ/mol) (**Fig 1d**). The energy requirement for the separation is significantly lower than suggested by ethanol-water separation models currently adopted in the CO2R processes TEAs (dashed line in Fig 1d). The concentrated ethanol stream can be used for direct conversion of ethanol to ethylene oxide[21], or other chemicals through conversion to ethylene[22].

The discovery of this energy efficient separation technique suggests the re-examination of the scalability of other electrocatalytic systems which were formerly considered as non-viable. We encourage the electrolysis community to consider embedding membrane separators in their systems, both experimentally and within the modelling studies. The latter will strongly benefit from a more detailed consideration of thermodynamic properties of the catholyte, which based on our results have a drastic effect on performance of the electrolysis-based chemical plant. This inspires further research into separation of other liquid products (e.g., acetate), as well as opens new opportunities for CO2R process optimization.

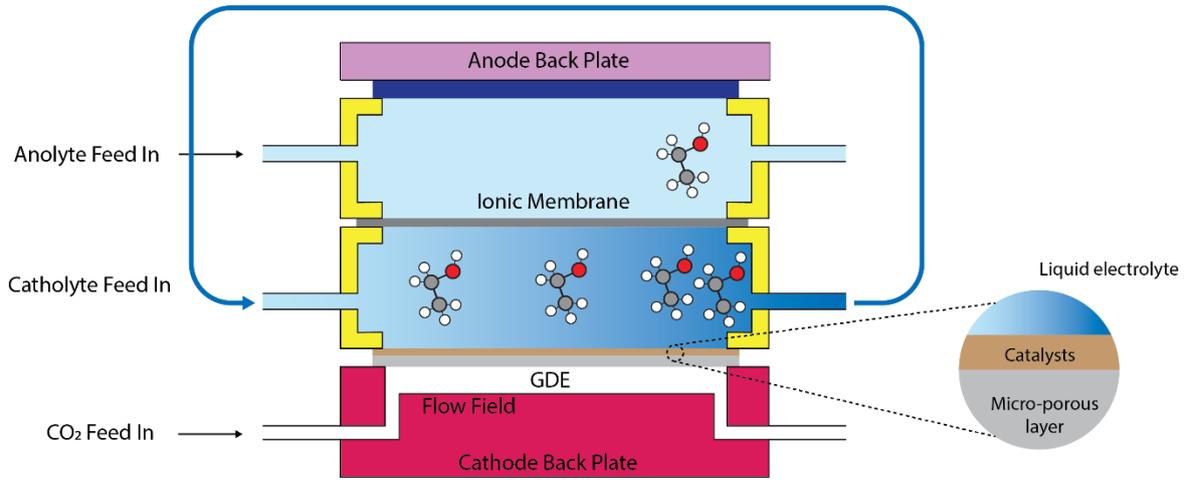

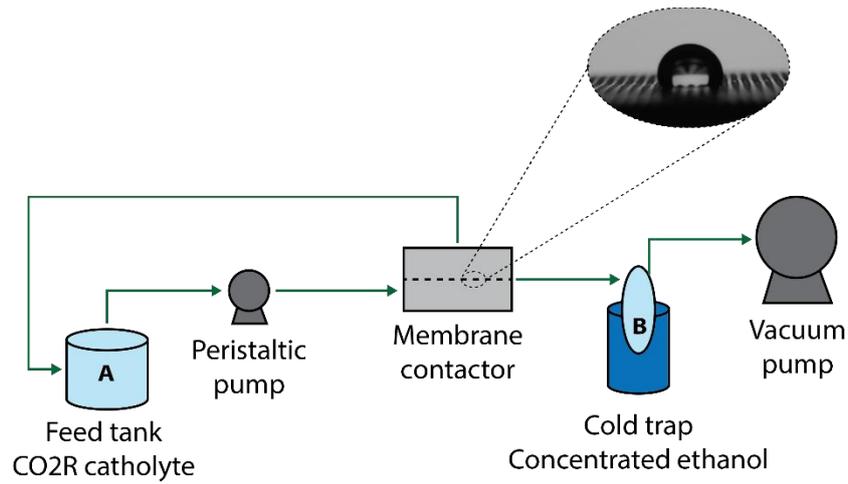

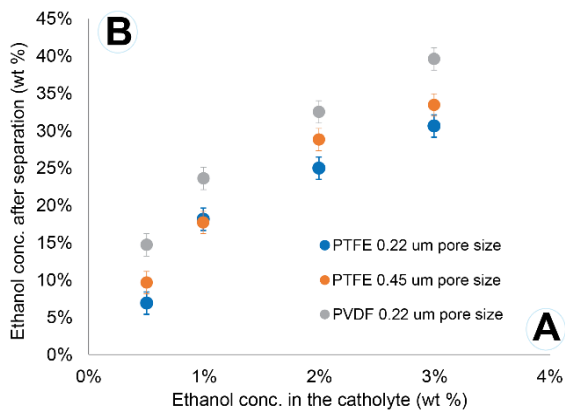

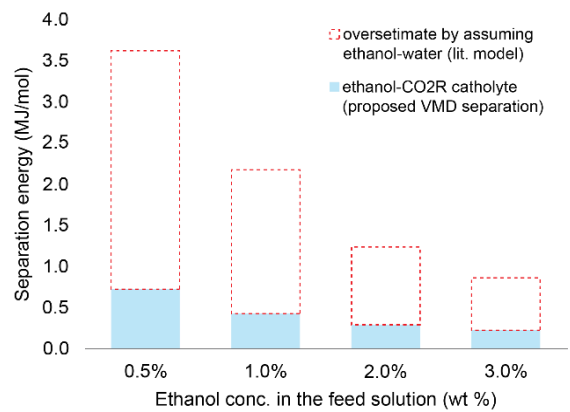

**Fig.1. Integration of ethanol separation by Vacuum Membrane Distillation (VMD) into CO2R**: a) cross-section of a gas-diffusion electrode based reactor, most commonly used for $CO_2$ electrolysis experiments under a high, industrially relevant current density; b) experimental setup used for the VMD experiments; details of the procedure and the equipment used are given in the SI, Experimental methods and Table S1, indexes A and B refer to the measurements plotted in 1.c., c) results of VMD runs performed for the separation of ethanol from 3.5 M KOH catholyte using PTFE and PDVF membranes of different pore sizes, under 25°C, 20-25 mbar pressure, d) energy requirement for the separation of ethanol from catholyte solution, 3.5 M KOH, calculated using the currently adopted approaches and proposed VMD technique.

**Acknowledgements**

The authors acknowledge the support of the National Research Foundation (NRF), the Prime Minister's Office, Singapore, under its Campus for Research Excellence and Technological Enterprise (CREATE) Programme through the eCO2EP project, operated by the Cambridge Centre for Advanced Research and Education in Singapore (CARES) and the Berkeley Education Alliance for Research in Singapore (BEARS). The contribution of Andres J. Sanz Guillen to discussions, proofreading and the enrichment of the visual content is gratefully acknowledged.

# Supplemental Information

# Energy efficient ethanol separation method for scalable CO$_2$ electrolysis

Magda H. Barecka,*[a,b,c], Pritika DS Dameni[c,d], Marsha Zakir Muhamad [c], Joel W. Ager *[e,f,g] and Alexei A. Lapkin [c,h]

i. Department of Chemical Engineering, Northeastern University, 360 Huntington Avenue, 02215 Boston, USA; email: m.barecka@northeastern.edu

j. Department of Chemistry and Chemical Biology, Northeastern University, 360 Huntington Avenue, 02215 Boston, USA

k. Cambridge Centre for Advanced Research and Education in Singapore, CARES Ltd. 1 CREATE Way, CREATE Tower #05-05, 138602 Singapore

l. Nanyang Technological University, Department of Mechanical Engineering, School of Materials Science and Engineering, 50 Nanyang Ave, 639798 Singapore

m. Berkeley Educational Alliance for Research in Singapore (BEARS), Ltd., 1 CREATE Way, 138602, Singapore

n. Department of Materials Science and Engineering, University of California at Berkeley, Berkeley, California 94720, USA

o. Materials Sciences Division, Lawrence Berkeley National Laboratory, Berkeley, California 94720, United States.

p. Department of Chemical Engineering and Biotechnology, University of Cambridge, Cambridge CB3 0AS, UK

**Supplemental text**

**Figures S1-3**

**Table 1-2**

**Supplemental references**

**Supplemental text**

*Experimental procedure: Materials*

Membranes used for separation were PTFE 0.22 μm and 0.45 μm pore size with PP reinforcement net and PVDF with 0.22 μm pore size, supplied as a courtesy by GVS Filter Technology. Membranes were cut into discs of 10 cm$^2$ diameter. Chemicals used for preparation of simulated products streams were purchased from Supelco (Ethanol 99%) or Sigma-Aldrich (Potassium hydroxide, Formic Acid, Acetic Acid) and used as obtained. Each of the simulated mixtures was mixed for 2 h in a closed container before the start of an experiment.

*Vacuum Membrane distillation experiments*

The experimental set-up (Fig. 1b in the main manuscript) consisted of a feed solution tank holding between 250-300 ml of feed solution, peristaltic pump (LongerPump BT300-2J with pump head YZ2515X), membrane module fabricated from stainless steel, with inner diameter of 9.9 cm and 15.4 cm$^2$ of the active membrane area (Fig. S2), cold trap immersed in a Dewar, and the rotary evaporator (Buchi R-215) used as source of vacuum and vacuum control. The experiments were carried at room temperature (25°C), deploying flowrate of catholyte of 60 mL/min and setting the vacuum at 25 mbar with 5 mbar of acceptable pressure variation during an experiment. The feed was directed from the solution tank to the membrane module and the ethanol separated was subsequently condensed in the cold trap immersed in isopropanol solution ( -20 °C). After each experimental run, the trap mass was measured on an analytical balance to identify the mass of separated ethanol (defined as permeate flux) and a sample of product was analyzed by a HPLC (Agilent 1100) using a SUPELCOGEL™ C-610H column, with a mobile phase being 5mmol $H_2SO_4$ solution in DI water.

*Process model*

To assess the energy input necessary for the separation of ethanol produced by $CO_2$ electrolysis, we adopted the approach used by Greenblatt et al.[1] as a part of Life Cycle Assessment framework associated with $CO_2$ electrolysis. The minimum amount of energy necessary for separation of distillation ($\Delta H_d$) is calculated as a function of the enthalpy of vaporization and the concentration of ethanol in the gas phase[2], following Equation 1. $x_{v,i}$ are the vapor phase mole fractions of ethanol (index 1) and solvent (water or catholyte; index 2); $H_{v,i}$ are the enthalpies of vaporization, and *r* is the reflux ratio, which can be approximated by $1/x_{v,1}$.

$$\Delta H_d = r(x_{v,1}\Delta H_{v,2} + x_{v,2}\Delta H_{v,2}) \tag{1}$$

For the ethanol-water solutions, literature reports were used to estimate the vapor phase fractions of ethanol corresponding to the given composition of the liquid phase mixture[3], and the enthalpies of evaporation. For the ethanol-catholyte solutions, the vapor phase fractions of ethanol corresponding to the given composition of the liquid phase mixture were derived from our experiments (Fig. 1c in the main manuscript). The enthalpy of vaporization of ethanol/water is affected by the respective boiling point, however, in the range of concentrations and respective boiling points that we are considering, these differences are minor. Thus, similarly to the approach adopted by Greenblatt et al.[1], the heat of evaporation was assumed as 46 kJ/mol (both for water and ethanol).

Equation 1 allows to assess only the energy necessary for vaporization, and not the energy necessary to heat-up the mixture to the boiling point (in case where thermal energy is used to drive the separation). While this assessment provides a simple basis for comparison of technologies and of catholytes, further studies will benefit from development of more detailed models, based on the experimental results reported in Fig. 1c.

a)

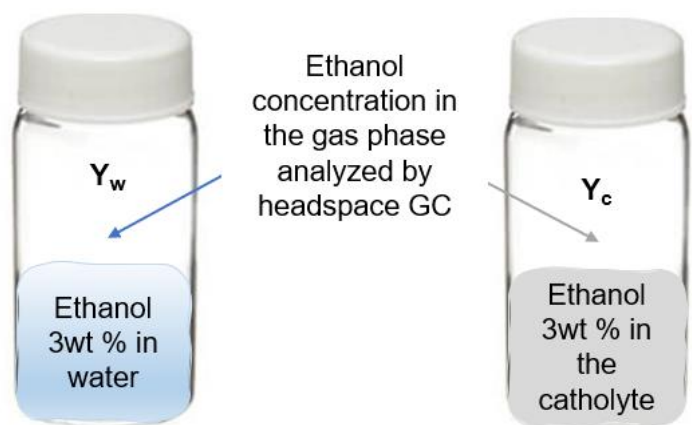

b)

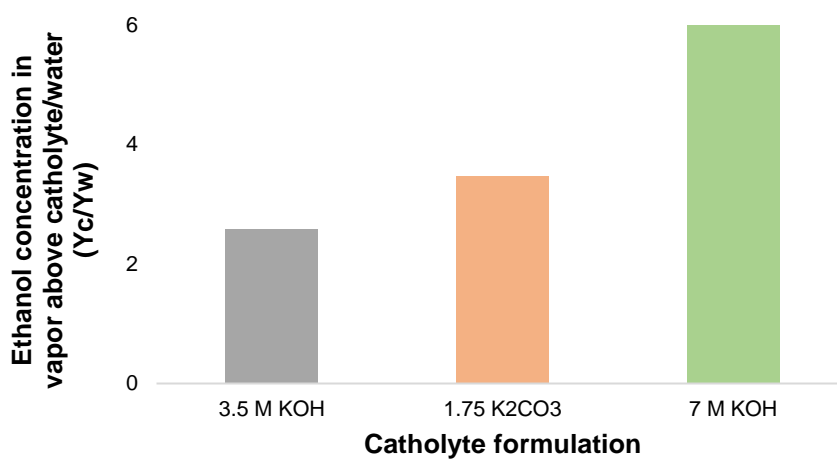

**Figure S1. Quantification of Vapour-Liquid Equilibrium for catholyte solutions used in CO2R. a)** A schematic representation of the experiment: we prepared a series of solutions that represent potential compositions of the catholyte during an electrolysis run, which is initially a pure hydroxide solution and gradually enriches in the captured $CO_2$: 3.5 M potassium hydroxide (KOH), 3.5 M potassium bicarbonate ($KHCO_3$), 1.75 potassium carbonate ($K_2CO_3$), 7 M KOH (as some studies deploy highly concentrated catholytes), and DI water (for comparison). To each solution, we added 3 wt.% ethanol. The solutions were transferred to gas chromatography vials, and kept for 2 h at 60 °C, before sampling of the gas headspace for analysis. Gas chromatography system used: GC Turbo Matrix HS40, FFAP column. FID detector, **b)** Results of the gas chromatography experiment are represented as a ratio of concentration of the gas phase above a catholyte solution to the concentration of the gas phase above water to highlight the enrichment. This measurement highlights the drastic different between the behaviour of ethanol-catholyte vs. ethanol water systems, the latter being typically considered in the CO2R literature. **Raw data for GC measurements are given in Table S1.**

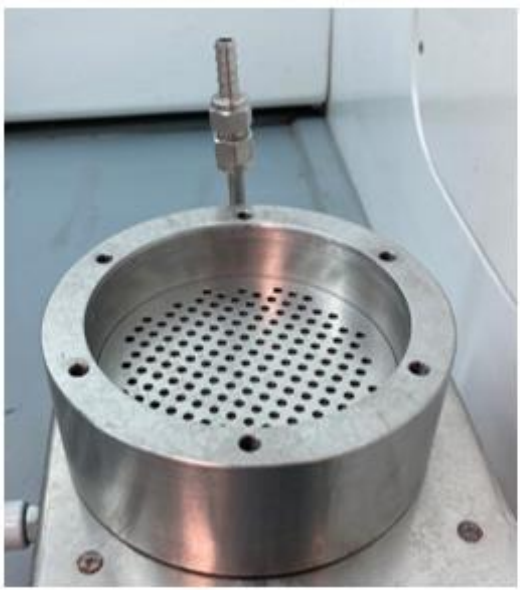 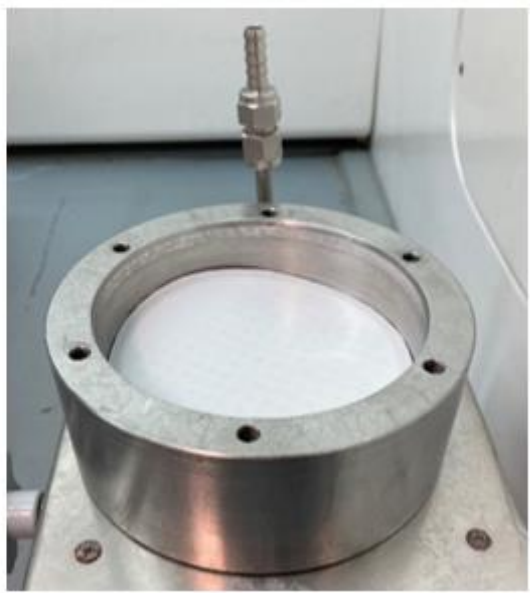

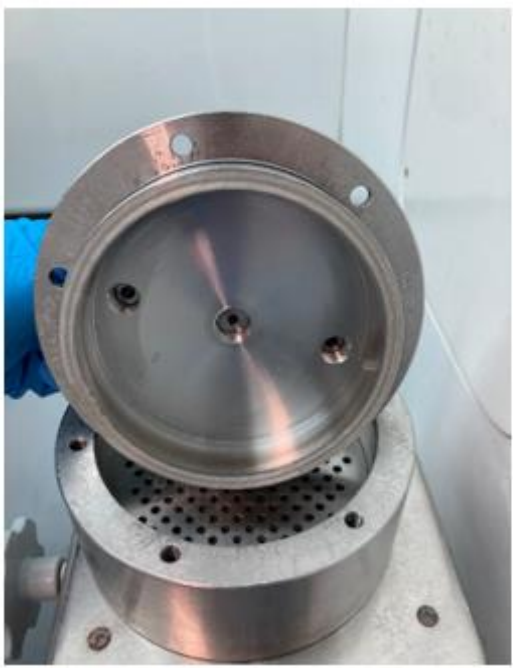 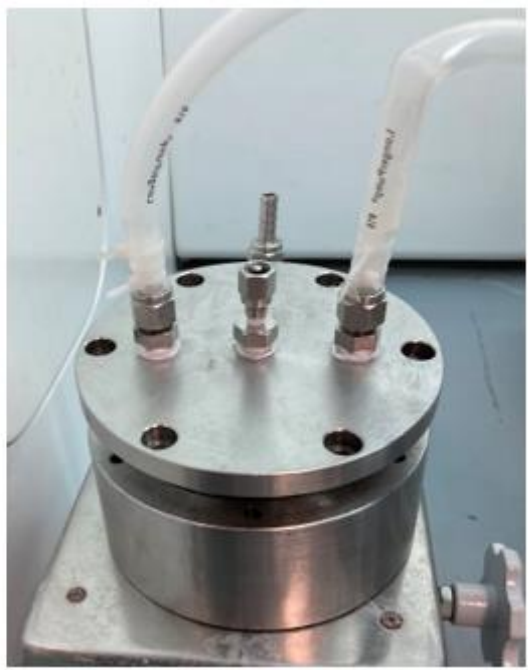

**Figure S2.** The vacuum membrane distillation cell used for the experiments, including a) stainless steel bottom plate, b) bottom plate with the membrane inside, c) upper plate with custom 3D printed polypropylene O-ring with square cross section, used to enhance sealing, necessary for separation under low pressures, d) view of the entire membrane holder with pump connections.

**Table S1.** Raw data recorded from headspace gas chromatography experiments described in Fig.S1. All samples were kept at 60°C for two hours before sampling.

| Sample type | Ethanol peak area |
|---|---|
| Ethanol 3wt% in water | 3347242 |
| Ethanol 3wt% 3.5 M KOH | 8653312 |
| Ethanol 3wt% 1.75 $K_2CO_3$ | 11659154 |
| Ethanol 3wt% 7 M KOH | 20819907 |

**Table S2. Detailed composition of catholyte streams used in vacuum membrane distillation experiments.** The majority of reports on $CO_2$ electrolysis cite only the Faradaic efficiency (FE) towards ethanol; however, from the perspective of product separation, mass concentration of ethanol is the key process design parameter. We therefore sought to assess the most representative range of mass concentration of ethanol obtained by electrolysis experiments (both for three compartment GDE cells[4–7] and tandem systems[8]), and recalculated Faradaic efficiencies to mass concentrations using a recent protocol[9], considering the catholyte volume, deployed current and experiment time. As a result, we assessed that ethanol can be concentrated up to 0.5 – 3 wt%, which is in a good agreement with an experimental report citing the actual ethanol mass concentration[10] and our experimental results. Besides ethanol, other liquid electrolysis products are always formed, so we considered them in our study (though only ethanol is being separated in our experiments).

| Compounds | Typical range of Faradaic Efficiencies | Resulting mass composition of the catholyte stream (wt%) | | | |
|---|---|---|---|---|---|
| **Ethanol concentration** | **< 20%** | **0.5** | **1.0** | **2.0** | **3.0** |
| Acetate concentration | < 20% | 1.3 | 2.6 | 5.2 | 7.8 |
| Formate concentration | < 7.5% | 1.5 | 3.0 | 6.0 | 9.0 |
| Potassium hydroxide solution concentration in all experiments | 3.5 M KOH | | | | |